\documentclass[10pt]{article}
\pdfoutput=1
\usepackage{graphicx}
\usepackage{amssymb}
\usepackage{epstopdf}
\usepackage{color}
\usepackage{amsmath}
\usepackage{amsfonts}
\usepackage{authblk}
\usepackage{lmodern}
\usepackage{vmargin}
\setmarginsrb{3cm}{2cm}{2.7cm}{2cm}{0cm}{1.5cm}{1.5cm}{1.5cm}

\title{Competitive allocation of resources on a network: an agent-based model of air companies competing for the best routes}
\author[1,2]{G\'erald Gurtner}
\author[1,3]{Luca Valori}
\author[1,4]{Fabrizio Lillo}
\affil[1]{Scuola Normale Superiore di Pisa, Piazza dei Cavalieri 7, 56126 Pisa, Italy}
\affil[2]{Deep Blue s.r.l., Piazza Buenos Aires 20, Rome, Italy.}
\affil[3]{Complex System Community (CSC), Università degli Studi di Siena, Via Roma 57, Siena, Italy}
\affil[4]{Santa Fe Institute, 1399 Hyde Park Road, Santa Fe NM 87501, USA}

\begin{document}
\maketitle
\begin{abstract}
We present a stylized model of the allocation of resources on a network. By considering as a concrete example the network of sectors of the airspace, where each node is a sector characterized by a maximal number of simultaneously present aircraft, we consider the problem of air companies competing for the allocation of the airspace. Each company is characterized by a cost function, weighting differently punctuality and length of the flight. We consider the model in the presence of pure and mixed populations of types of airline companies and we study how the equilibria depends on the characteristics of the network.

\end{abstract}



\section{Introduction}

Networks represent the natural structure to describe the dynamics of many complex systems \cite{boccaletti,caldarelli}. Examples include the dynamics of epidemiology, information propagation on the Internet, percolation, opinion spreading, systemic risk, etc. Transportation and traffic \cite{helbing} is another field where networks describe naturally the structure and the dynamics of the system. Recently, an increasing attention is being devoted to the network description of transport systems and of transport dynamics on networks. This is partly due to the fact that transport systems, along with other infrastructures like power grids or communication networks, are fundamental elements of our societies and economies.  The description of these infrastructures as networks provides the link between the interaction of isolated agents and the resulting macroscopic phenomena. Moreover, it allows to make a distinction between the infrastructure (e.g. the road network) over which policy makers have typically some degree of control, and the traffic resulting of the choice of different agents conditioned on the infrastructure.

Transportation needs are increasing very rapidly and this is creating new challenges to policy makers, because transport infrastructures are a scarce and costly resource. The increasing traffic risks to overload the infrastructure system, enhancing the probability of congestions, incidents, and strongly suboptimal use of the infrastructures. From the point of view of the users (e.g. transportation companies) this increases the effort required to find better strategies, whose success depends, among other things, on the strategies adopted by the other users. 

A paradigmatic example, considered in this paper, is the air transportation \cite{li,guimera,colizza,lacasa,zanin,cardillo,plos}; for a recent review see \cite{zaninlillo}. Air transport is increasing worldwide at a very fast pace, estimated at 5\% per year in US and Europe. Policy makers are aware of the fact that the current system will be at its capacity limits in a few years because of the increase of traffic demand and new business challenges. For this reason, large investment programs like SESAR in Europe and SingleSky in the US have been launched.  Thus airspace is becoming a scarce resource, especially in congested situation, like, for example, during major shutdowns of large areas (extreme weather, strikes, volcano eruptions, etc.). 

The problem of occupation of the airspace can be seen as an allocation of resources in a network. In fact, as described for example in \cite{plos}, for what concerns air traffic control the airspace can be represented as a network of sectors.  Sectors are the smallest operational pieces of airspace and as such are controlled by a pair of controllers. One in particular is in charge of the interaction with the other neighboring sectors. The network is built by associating to each sector a node and two nodes are connected if they share a common boundary. This procedure defines the {\it sector network} and a flight can be represented as a walk on this graph. Each node (sector) is characterized by a capacity, which roughly speaking is the largest number of aircraft that can be simultaneously present in the sector, and a crossing time, which is the typical time needed for an aircraft to cross the sector. When traffic increases, overloaded sectors could appear. For this reason the airspace is currently regulated by a Network Manager (NM). Air companies submit hours - or a day - before the departure of a flight a flight plan to the NM, containing the information on the desired route and time of the flight. The NM checks if some sector would exceed its capacity and if this is the case it asks the airline to submit another proposal, until a flight plan without conflict is accepted. The strategy of the airline company depends on its business model (e.g. if it is a low cost or a major carrier). Its success in achieving the desired slots depends on the infrastructure (the topology of the sector network, the airport location, the available departing times, etc) {\it and} on the strategies followed by the other airline companies.

In this paper we present a toy, yet phenomenologically rich, model of this allocation of resources on the airspace, by modeling it as a sector network and by considering different strategies of the airline companies. We want to study how the topology of the network and the structure of the departing times affect the satisfaction (i.e. the fitness) of a specific type of company. We consider first the case when only one type of company is present. However the most interesting case is when a mixed population of companies is simultaneously present. In this case we find that the satisfaction of a company depends on the mixing composition, i.e. on the fraction of other airlines following the same strategy. We show how all these factors explain the satisfaction of different types of company. Some of these factors (the network topology and the departing times) can be regulated externally by the policy maker, while others (the mixing composition) depend on the airline population and market forces.

The paper is organized as follows. In Section \ref{sec:alloc} we present the current procedure for flight allocation in the airspace. In Section \ref{sec:model} we present our model and in Section \ref{sec:results} we present the results. 
Finally in Section \ref{sec:conclusions} we draw some conclusions. 

\section{The allocation of air traffic routes}\label{sec:alloc}

\begin{figure}[t]
\begin{center}
\includegraphics[width=0.43\textwidth]{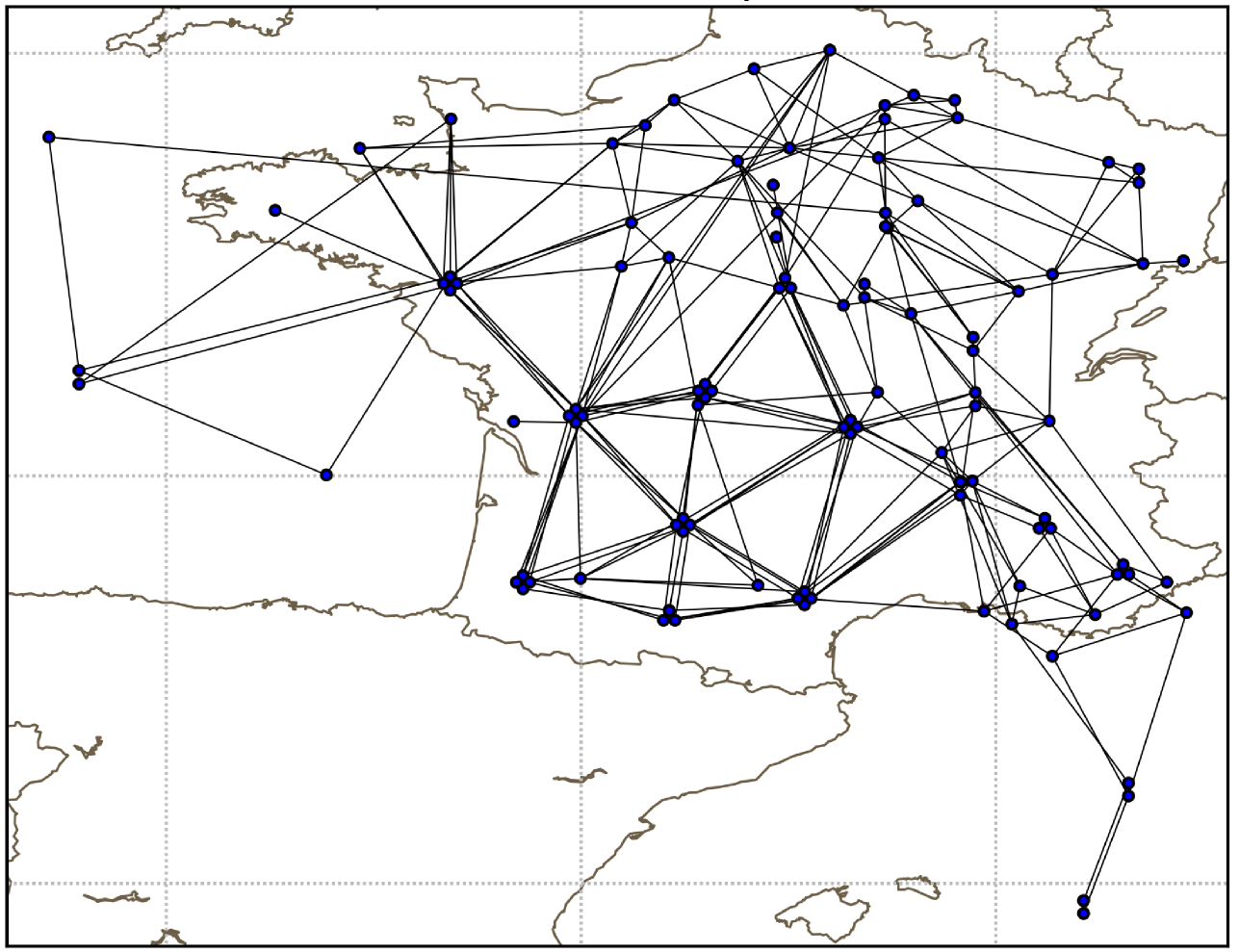}
\includegraphics[width=0.53\textwidth, height=5cm]{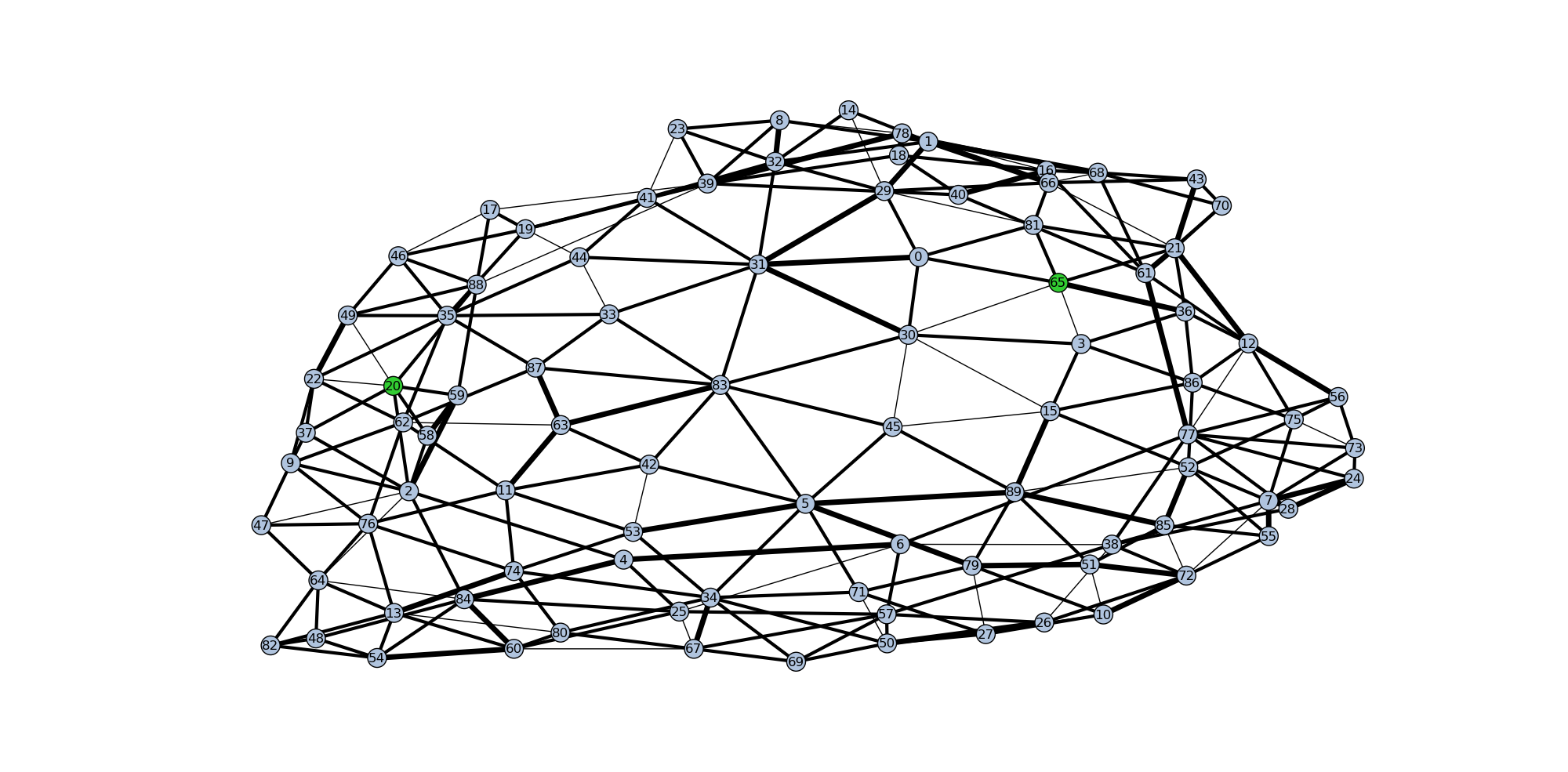}
\end{center}
\caption{Left panel. Network of sectors of France. Each circle represents a sector and is situated at the geometrical centroid of it. The sectors on top of each other are slightly shifted a bit to better see them. Right panel. 
Graph used for running the model. The graph is a Delaunay triangulation of 90 randomly chosen points on a plane. Green nodes represent sectors with airports. The thicker is the edge, the higher is the associated weight, i.e. the number of flights crossing the edge. Note that this representation of the graph is purely topological.}
\label{fig:sectorNetwork}
\end{figure}

In order to understand our model, it might be helpful to summarize what is the real current procedure used by the Airline Operators\footnote{We will use the terms air companies and air operators interchangeably.} (AO) to allocate their flights. A flight plan is a trajectory connecting the departing and the arriving airport. Simplifying a bit, a flight plan contains the information on the position of the aircraft at each time. The airspace is partitioned into sectors, which are portions of space and the elementary units of air traffic control. The left panel of Figure \ref{fig:sectorNetwork} shows the network sectors of France (see \cite{plos} for more details). In the network of sectors each sector is a node and two nodes are connected if at least one flight goes directly from one node to the other in the considered time interval. At a given altitude the graph is very close to a planar one. Deviations from planarity are observed because the partition of space is not uniform at different altitudes. For simplicity in our model we will use a representation of the sector network as a planar graph.

Each sector has a capacity, which is roughly defined as the maximal number of flights that can be simultaneously present in it. For each flight, a flight plan has to be submitted and approved by the Network Manager (NM). This is a regulatory entity monitoring the planned occupation of the sectors and controls if a newly submitted flight plan violates some constraint on the maximal capacity of a sector. Specifically, after having collected information about weather, number of passengers, connecting flights, etc.,  the AO computes the best flight plan taking into account fuel costs and delays, and submits it to the NM. If the flight plan does not introduce any sector capacity overload, the flight plan is accepted and the NM updates the load of the sectors crossed by the flight, taking into account the entry and exit time for each sector. If the flight plan is rejected, the AO computes the next best trajectory and submits it to the NM. This process continues until a flight plan without conflicts is accepted. The ranking of the possible flight plans for a given flight depends on the strategic choices and the utility function of the AO.  Notice that when submitting a flight plan to the NM, the AO does not know the sector load or what the other AOs did or plan to do.

\section{The model}\label{sec:model}

The model is composed by the airspace, modeled as a network of sectors, and the AOs who have a cost function and submit sequentially flight plans to the NM. We describe them separately below and we then introduce the metrics we use to characterize the behavior of the model.

\subsection{The airspace}

We generate artificial networks of sectors by performing a Delaunay triangulation \cite{delaunay}. Specifically, we generate a set of uniformly distributed points in a square region of the plane (i.e. between -1 and +1 both in x and in y). Then we perform the Voronoi tessellation starting from these points. In such a tessellation, the points inside the boundaries of each tile are closer to its centroid point than to any other point in the starting set. In particular the cells are convex. These tiles represent airspace sectors. The Delaunay triangulation is obtained by plotting the dual graph associated with the Voronoi tessellation.  The Delaunay triangulation is a planar graph and each vertex has on average six surrounding triangles (i.e. the mean degree is equal to 6, which is also the maximum average degree for planar graph). 

Each point of the network represents a sector (which might contain an airport). To each sector we associate a capacity, which is the maximum number of flights simultaneously present in the sector\footnote{The operational definition of sector capacity is the maximal allowed number of flights which, in a given hour, has traveled through the sector (including those who are entering and those which were already there at the beginning of the hour). However our investigation on empirical data on the European airspace (unpublished) indicates that the two measures are approximately proportional one to each other.}. For simplicity in the model we assign the same capacity (equal to $5$) to all sectors.

In our model, time is a continuous variable. Hence, the crossing times between sectors are real numbers that we choose proportionally to the Euclidean distance between the centroids of the sectors.  The constant of proportionality is equal to the inverse average speed of aircraft. 

Finally we choose a set of airports randomly, with the only constraint that the source and the destination of a flight cannot be closer than a minimal topological distance. The analysis presented in this paper focuses on the simplest case of two airports.
The right panel of Figure \ref{fig:sectorNetwork} shows the Delaunay triangulation of 90 randomly chosen points on a plane, used in the first part of the paper to test the model. The figure also shows the airport nodes. In the following sections we will show the results with this choice. In Section \ref{sec:topology} we will discuss how the choice of the airports affects the model's results. 

\subsection{Agents}

The agents of the model are the Airline Operators (AOs), who try to get the best trajectories for their flights. This best trajectory has two components: the geometrical length and the times of departure/arrival. Different AOs, which may have different goals, are competing one with each other for the ``best'' slots and trajectories, based on business considerations and constrained by the structure of the airspace. As described above, the network manager tries to fill the airspace as best as possible, its main concern being to avoid to overload the airspace, in order to guarantee safety \cite{conway}. In our model, as well in reality, the network manager is very passive, since it takes only propositions from the AOs and tries to fill the airspace, without making counterpropositions. 

In detail, for each flight an AO chooses a departing and arrival airport, a desired departing time, $t_0$, and selects a number $N_{fp}$ of flight plans. These are the best ones according to a cost function taking into account the length and the punctuality of the flight. Specifically, the $k-$th flight plan, $k=1,...,N_{fp}$, is defined as a pair $(t_0^k, {\bf p}^k)$, where $t_0^k$ is the time of departure and ${\bf p}^k$ is an ordered set containing the sequence of sectors followed by the aircraft. Each AO has a cost function for its flights of the form
\begin{equation}
c(t_0^k,{\bf{p}}^k) = \alpha {\cal L}({\bf{p}}^k) + \beta (t_0^k-t_0),
\end{equation}
where ${\cal L}( {\bf{p}}^k)$ is the length of the path on the network (i.e. the sum of the lengths of the edges followed by the flight). We also assume that flights are only shifted ahead in time ($t_0^k \ge t_0$) by an integer multiple of a parameter $\tau$ which is taken here as unitary. The parameters $\alpha$ and $\beta$ define the main characteristics of the company. A high value of $\beta$ simulates a company eager to have its flights on time even if with a higher fuel cost. On the other hand, a high $\alpha$ simulates a company more preoccupied by the length of the trajectory. In the following, we use often the two extreme configurations: $\beta/\alpha\gg 1$, in which the company cares mostly about punctuality and takes any path that guarantees the desired departing time, and $\beta/\alpha\ll 1$, for which the company cares mostly about the length of the path, thus taking the shortest one, possibly shifted in time. We will call the first type of company ``R'' (for rerouting) and the second one ``S'' (for shifting). From an operational point of view the S companies mimic the low-cost companies, while the R companies mimic the major airlines.


A key parameter of the model is the desired departing time $t_0$ chosen by the AO. Similarly to what is observed in real airports, we assume that departing times are drawn from a distribution inside the day characterized by a certain number of peaks or waves\footnote{In reality departing time is the result of a complex optimization taking into account many operational and economic aspects, such as the availability of aircraft, crew, the demand from customers, etc. Clearly our toy model does not describe these aspects.}. 
We define first $d$ as the length of the ``day'', i.e. the time window of departure for all flights. In this time window, we define $N_p$ peaks of length 1 (typical crossing time of a sector), by setting a time $\Delta t$ between the end of the peak and the beginning of the next one (thus, $N_p = d/(\Delta t+1 )$). The parameter $d$ is fixed to $24\tau=24$ (notice that this number not necessarily must be interpreted as one day). Then, we define either a total number of flights and divide them equally between peaks (within which they depart at random), or we set a given number of flights per peak.

The dynamics of the model is as follows. At each time step, one AO, characterized by the $\alpha$ and $\beta$ parameters, is selected in a random way from the population of AOs (see below for the different choices of the set). The AO chooses the departing and arrival airport and the desired departing time $t_0$ drawing it from the departing time distribution. It computes the $N_{fp}$ best flight plans for the flight according to its cost function and submits them, one by one in increasing order of cost, to the NM.  The NM checks if the flight plan creates an overload in any of the crossed sectors. If this does not happen, the flight plan is accepted and the NM updates the load of all the crossed sectors. Instead, if there is at least one overloaded sector, the NM rejects the flight plan, and the AO submits the next best flight plan for the same flight, according to the ranking created by its cost function. This process continues until one flight plan is accepted or all the $N_{fp}$ flight plans are rejected.

\subsection{Metrics}

We describe here the metrics we use to measure the performance of the AOs and of the whole system. These quantities are  related to the cost. Specifically, for each flight $f$, we define the satisfaction of the AO as 
\begin{equation}
{\cal S}_f=c_f^{best}/c_f^{accepted},
\end{equation}
where $c_f^{best}$ is the cost of the optimal flight plan for the flight $f$ according to the AO cost function (this flight is also the first flight plan to be submitted for the flight), and $c_f^{accepted}$ is the cost of the flight plan eventually accepted for this flight. If no flight plan has been accepted, we set ${\cal S}_f$ to 0. Note that ${\cal S}_f$ is always between 0 and 1. The value 1 is obtained when the best flight plan is accepted.

The satisfaction is a metric defined for a single flight. We also define the average satisfaction across all the flights of AO identified by the label $i$ as
\begin{equation}
{\cal S}^{(i)}=\frac{1}{M^{(i)}}\sum_{f=1}^{M^{(i)}} {\cal S}_f^{(i)},
\end{equation}
where $M^{(i)}$ is the number of flights of the AO and ${\cal S}_f^{(i)}$ is the satisfaction of the $f$-th flight of  company $i$.

When more than one AO is present in the system, the definition of the satisfaction for the whole system is not unique and more complex.
Let us consider the simple case of two air companies, $AO_1$ and $AO_2$, and assume that there is a fraction $f_1$ of flights belonging to company $AO_1$ and a fraction $f_2$ of flights belonging to company $AO_2$. Finally, let  ${\cal S}^{(1)}$ and ${\cal S}^{(2)}$ be the average satisfaction of the two AOs.

The global satisfaction is the sum of the weighted satisfactions of the different air companies, where the weights are given by the corresponding portions of flights, i.e. 
\begin{equation}\label{eq:GS}
{\cal S}^{TOT} = f_1 \times {\cal S}^{(1)} + f_2 \times {\cal S}^{(2)}.\\
\end{equation}
Thus ${\cal S}^{TOT}$ is the average satisfaction of the whole population.

In the normalized global satisfaction, instead, we consider the satisfaction of an AO by normalizing it over the one computed when the other company is not present.  
\begin{equation}\label{ngs}
\tilde{\cal S}^{TOT} =  f_1 \times \frac{{\cal S}^{(1)}}{{\cal S}^{(1)}(f_2=0)} +f_2 \times \frac{{\cal S}^{(2)}}{{\cal S}^{(2)}(f_1=0)}. 
\end{equation} 
$\tilde{\cal S}^{TOT}$ is an interesting measure because it is insensitive to the different level of satisfaction observed when each AO is alone.

\section{Results}\label{sec:results}

In this Section we present the results of our model. We first fix the topology of the network and the location of the two airports (see the right panel of Figure \ref{fig:sectorNetwork}) and we consider separately the case of a pure population of AOs, all characterized by the same cost function, and the case of a mixed population. In the latter case, we have two types of diametrically opposed AOs, one of type R and one of type S. We then consider the role of network topology and departing time pattern in the competitive advantage of the two types of AOs.  

\subsection{Pure populations}

\begin{figure}[t]
\begin{center}
\includegraphics[width=0.48\textwidth]{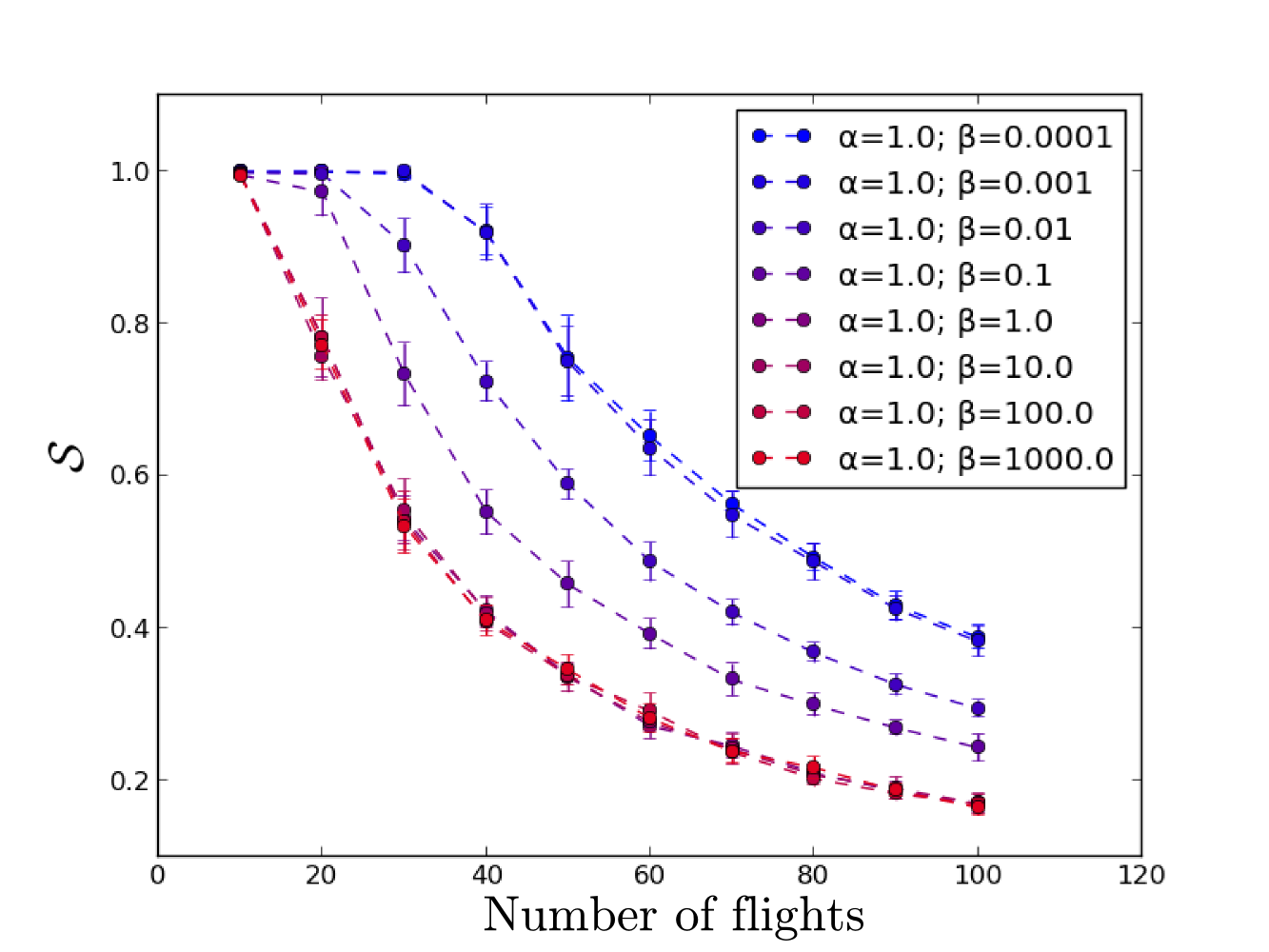}
\includegraphics[width=0.48\textwidth]{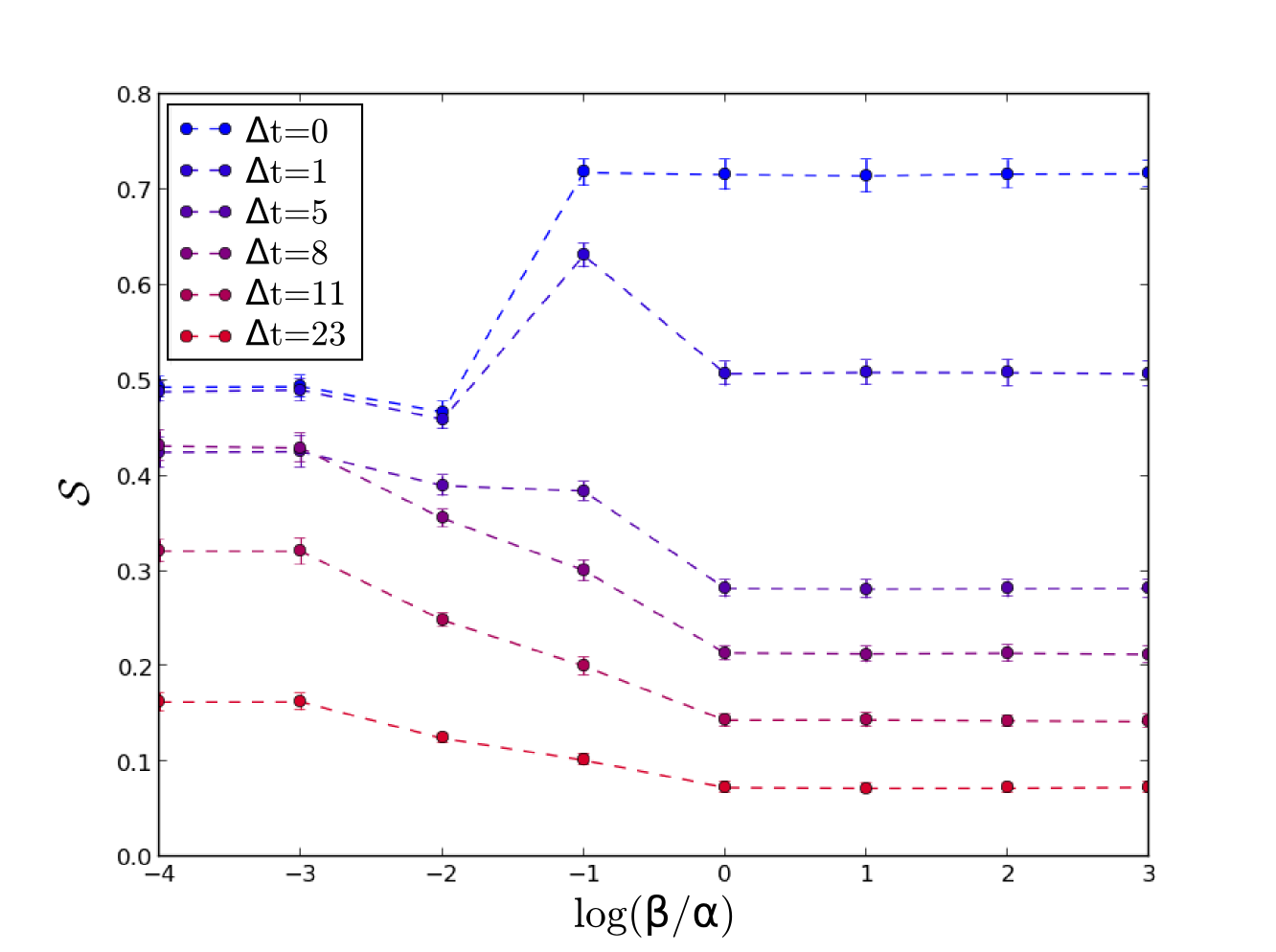}
\end{center}
\caption{Left panel. Satisfaction of an AO in the pure population case versus the total number of flights for different values of the ratio $\beta/\alpha$. Only one type of companies is considered and the desired departing pattern has only one peak ($\Delta t=23$). Right panel. Satisfaction as a function of the ratio $\beta/\alpha$ for different values of $\Delta t$. The simulation is done with a number of flights per peak kept fixed.}
\label{fig:pure_1}
\end{figure}

We consider first the case in which there is only one type of air company characterized by the cost function parameters $\alpha$ and $\beta$. This is the case of a {\it pure population}. We studied how the satisfaction of the company depends on the cost function, the total traffic, and the structure of departing times. The left panel of Figure \ref{fig:pure_1} shows how the satisfaction of the AO varies with the number of flights, for different values of the ratio $\beta/\alpha$. In this figure we set $\Delta t=23$,  i.e. there is only one departing peak at $t=0$. This way we do not consider the effect that delayed flights of one wave have on the flights of the following wave. As expected, the satisfaction is monotonically decreasing with the total number of flights, since with low traffic there are less conflicts. For small values of $\beta/\alpha$ (S companies) there is a clear plateau for small values of  the number of flights. This corresponds to the case when there are no conflicts, because the traffic is low. Similarly to what observed in other traffic models \cite{Nagel}, below a threshold for the number of flights, the satisfaction is constant, while at the threshold there is a transition which is discontinuous in the first derivative. This transition is not observed for large values of $\beta/\alpha$ (R companies). This difference is due to the fact that for not too heavy traffic, the S companies can shift in time the departure without increasing the cost and conflicting with other aircraft. For R companies rerouting does not help much since some sector will be soon fully occupied avoiding to benefit from a alternative path. This explains also why generically S companies are more satisfied than R companies (when they are alone).

How does this behavior change when one considers different departing patterns? In the right panel of figure \ref{fig:pure_1}, we present the satisfaction against the ratio $\beta/\alpha$ for different values of $\Delta t$. Moreover we fix the number of flights per peak, hence the overall satisfaction is decreasing when $\Delta t$ decreases because there are more waves and thus more flights in total. For large values of $\Delta t$, i.e. few and well separated waves, the satisfaction decreases monotonically with the ratio $\beta/\alpha$. Company S is always doing better than company R. On the contrary, when there are many waves (small $\Delta t$), the satisfaction is increasing with the ratio, i.e. companies R are better off. This is because in this last situation the companies which shift in time (i.e. S) find other companies ahead: it is thus better to change the route instead. It is also interesting to see that there is an intermediate situation ($\Delta t=1$) for which none of the extremes is better. In other words companies doing a compromise between length of path and time of departure have a higher satisfaction.

In conclusion, when only one type of companies is present there is an interplay between the company with the greatest advantage and the structure of the departing times. There is no company that has in absolute term an advantage, but the fittest company depends on the environment, as described by the departing time pattern.

\subsection{Mixed populations}\label{sec:mixedpop}

Now we consider a system where there are two types of competing companies, each of them characterized by different parameters of the  cost functions. A fraction $f_S$ of the flights are of type S and with $\beta/\alpha=10^{-3}$. The remaining fraction $f_R=1-f_S$ are of type R and characterized by $\beta/\alpha=10^3$. 

\begin{figure}[t]
\begin{center}
\includegraphics[width=0.48\textwidth]{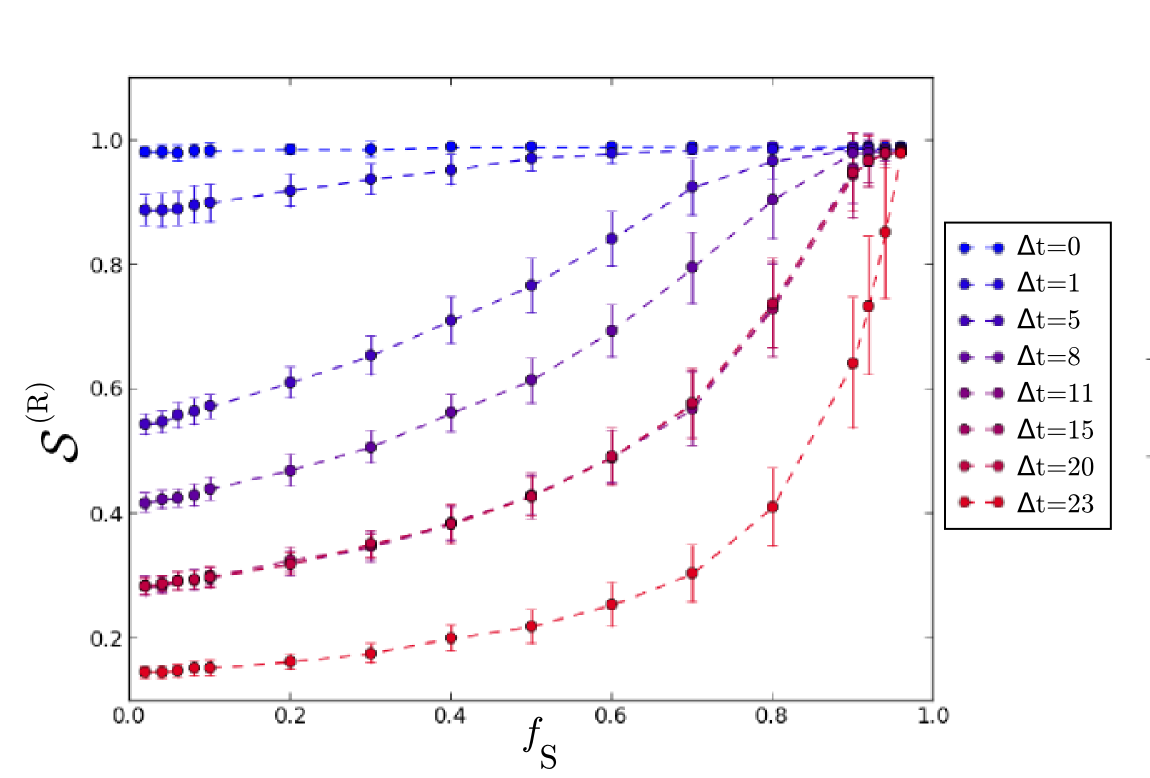}
\includegraphics[width=0.48\textwidth]{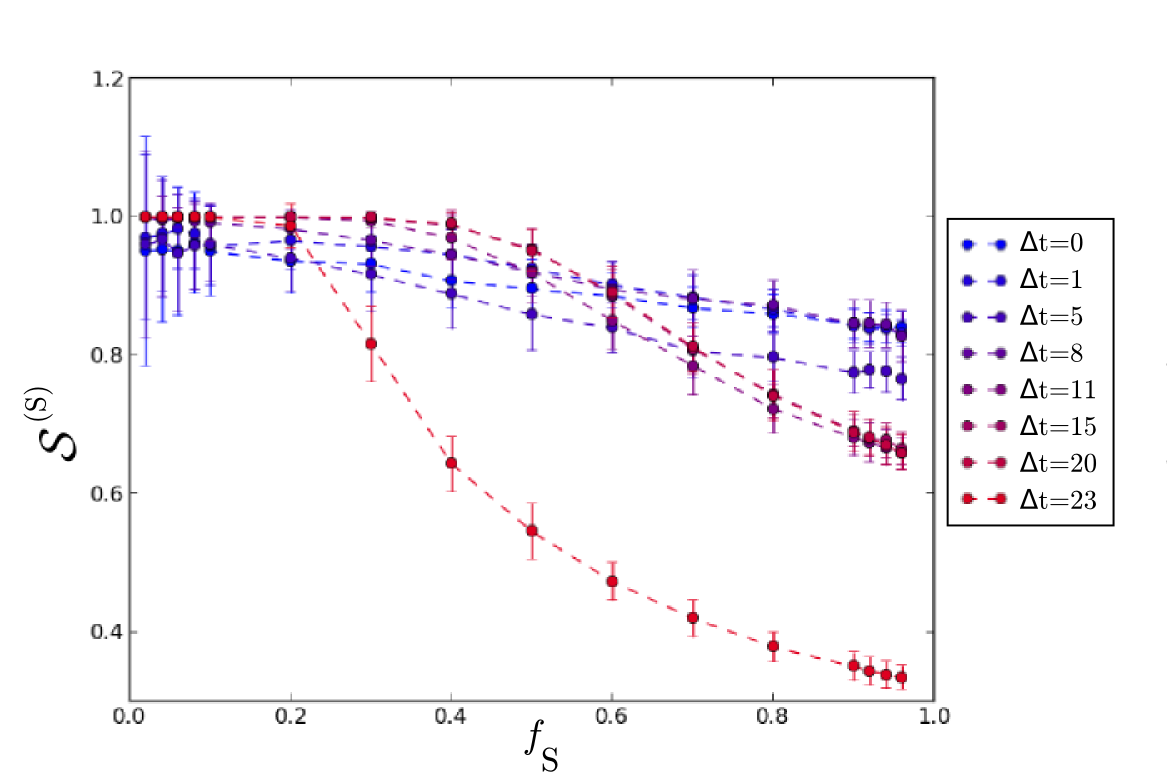}
\end{center}
\caption{Satisfaction of company R (left) and S (right) against the fraction $f_S$ of company S for different values of $\Delta t$ in the mixed population setting.}
\label{fig:mixed_1}
\end{figure}

In figure \ref{fig:mixed_1} we show how the satisfaction of each type of company depends on the mixing of the population.
As a general trend, we observe that the satisfaction of company R increases with the fraction $f_S$ of S companies, while the satisfaction of S companies decreases with it (and increases with $f_R=1-f_S$). This means that a company gets a higher satisfaction when it is surrounded by companies of the other type, rather than its own. We can understand this result by considering that if everybody wants to shift the flight plan, it is better for a company to reroute, because the suboptimal routes will be left free. Similarly, the satisfaction of S company is maximum when the company is alone since there will be free space shifting in time the flight, while at the planned time all the routes are jammed. However, the behavior of the satisfaction of company S is a bit more complex. The different curves corresponding to different values of $\Delta t$ are crossing each other. This means that for different values of $f_S$, it is better sometimes to have a small $\Delta t$, and sometimes a large $\Delta t$. For instance, for $f_S=0.4$, it is better to have $\Delta t=20$ rather than to have $\Delta t=1$, which is perfectly normal for S companies. On the other hand, for $f_S=0.8$, the contrary happens, and suddenly it is better for S companies to have a more uniform distribution of departing times. The whole picture is even more complicated by the fact that this behavior is not monotonic with $\Delta t$. With the same example, with $f_S=0.4$, it is better for S to have $\Delta t=20$ rather than 1, but it is much worse to have $\Delta t=23$. This non trivial effect of mixing different companies gives a rich behavior in terms of optimization of the total satisfaction. 

\begin{figure}[t]
\begin{center}
\includegraphics[width=0.48\textwidth]{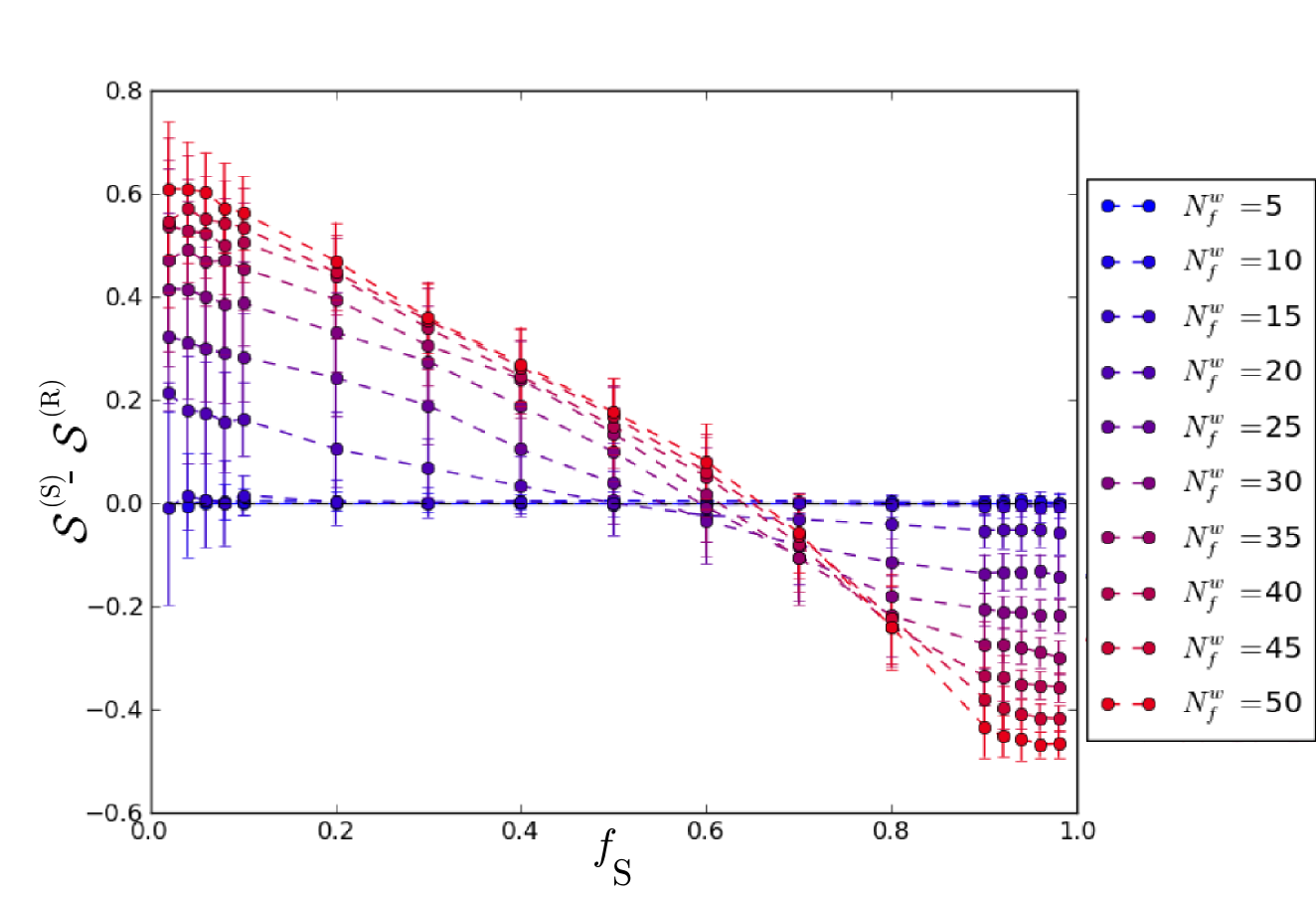}
\includegraphics[width=0.45\textwidth]{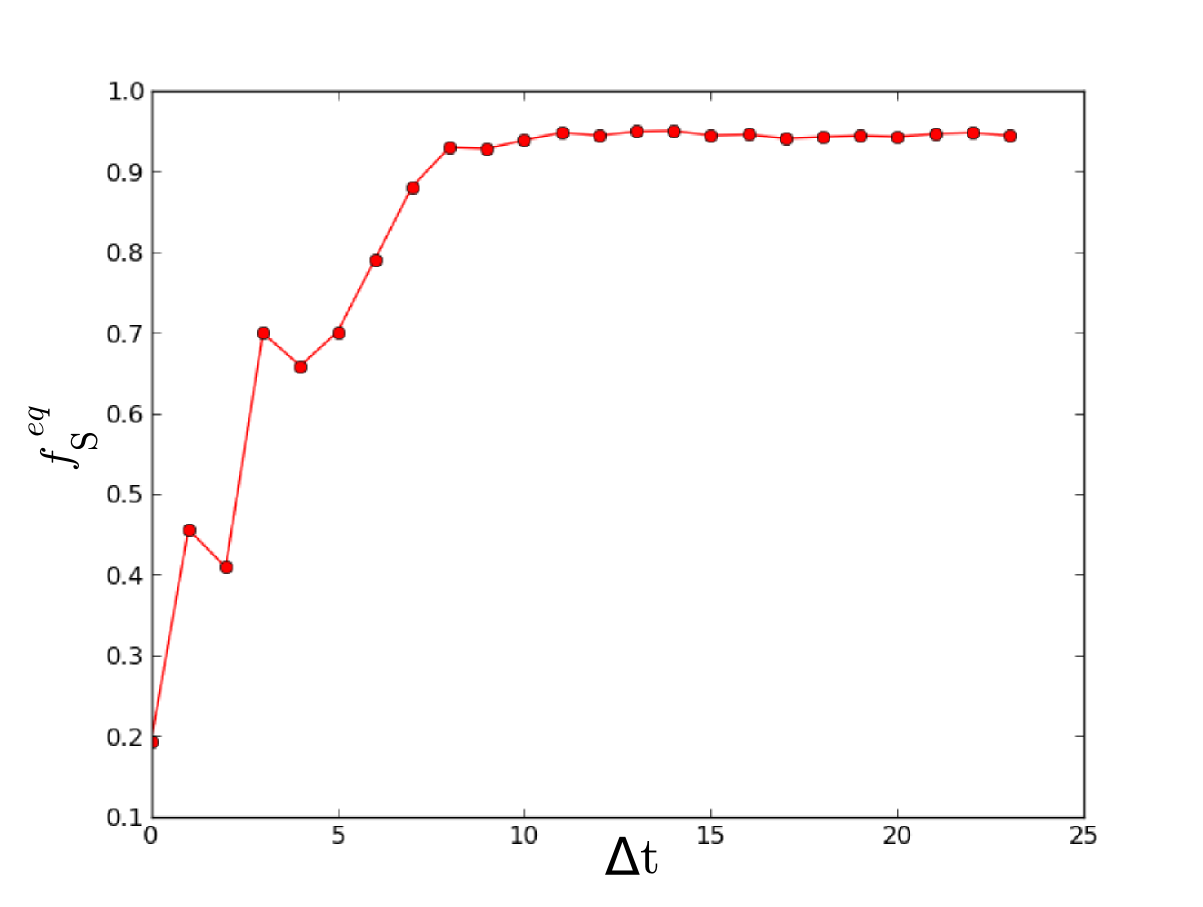}
\end{center}
\caption{Left panel. Difference between the satisfaction of S companies and R companies as a function of the fraction $f_S$ of S companies for $\Delta t= 5$. Right panel. Relation between the equilibrium mixing fraction, $f_S^{eq}$, as a function of the departing time pattern parameter $\Delta t$.}
\label{fig:mixed_2}
\end{figure}

The above picture is  clearer when we compare the satisfaction of the two different companies as a function of the mixing of the population.
The left panel of Figure \ref{fig:mixed_2} shows the difference between the satisfaction of S companies and R companies as a function of the fraction $f_S$ of S companies for $\Delta t= 5$. In this plot we consider different numbers of aircraft per wave, i.e. we change the total traffic. We see that the difference is positive (negative) when $f_S<f_S^{eq}$ ($f_S>f_S^{eq}$), where $f_S^{eq}\simeq 0.65$, i.e. broadly speaking when there is a minority (majority) of S companies. Note that the equilibrium fraction $f_S^{eq}$ where the difference in satisfaction is zero slightly depends on the total traffic. Moreover for a fixed $f_S$ the difference in satisfaction increases and then saturates with the traffic load.  

The equilibrium point $f_S^{eq}$ depends on the parameter $\Delta t$ characterizing the desired departure pattern. In the right panel of Figure \ref{fig:mixed_2} we show explicitly that this dependence is monotonically increasing. This means that when $\Delta t$ is small (large), S (R) companies have an advantage only if they represent a tiny fraction of the population. For intermediate values of $\Delta t$, the critical fraction is closer to $0.5$.

We now comment on the significance of these relations. The first observation is that the equilibrium between the two populations is related to the value of $\Delta t$ separating waves of departure. Thus regulating the structure of departures in airports (for example increasing their capacity) modifies the advantage of different companies and leads to another equilibrium point in their mixing. Second, from an evolutionary perspective, the difference in satisfaction is reflected in a different capability of each company to continue its business in the future. Even if we are not modeling the population dynamics of  the AOs, one could expect that a higher satisfaction translates into a larger possibility to expand business. This is the usual mathematical connection between evolutionary game theory and population dynamics \cite{Sigmund}.  In particular the behavior of the difference of the satisfaction of the two strategies is analogous to the one observed in some games, such as, for example, the  hawks and doves game \cite{Maynard,Sigmund}. 
 We can therefore conclude that a company which uses the pure strategy S with probability $f_S^{eq}$ and the pure strategy R with probability $1-f_S^{eq}$ cannot be invaded by other (possibly mixed) strategies. In terms of game theory such a strategy is an Evolutionary Stable Strategy. This strategy therefore will dominate the market. It is important to notice that, differently from the majority of papers on game theory, here we have derived the payoff structure, rather than postulating it from the beginning. It is a consequence of the structure of our model. Another important difference with respect to the typical game (e.g. the hawks and doves game) here the interaction between the agents (air companies) is not binary but is mediated by the network manager who allocates the flights, rejecting those that create a conflict. 

From another perspective, the behavior of our model is similar to the one observed in the minority game \cite{minoritybook}. In this model the payoff depends on the fraction of agents following one of two strategies, and the more minoritarian is the strategy that one chooses, the higher will be her payoff. This is analogous to what is observed in the left panel Figure \ref{fig:mixed_2}, with the difference that, while in minority game the equilibrium position corresponds to $f_S^{eq}=1/2$, here the equilibrium depends on the departing pattern. 
The strategy where companies choose to be an S company with probability $f_S^{eq}$ is the unique symmetric Nash equilibrium of the game. Again, the structure of the payoff is in our case an output of the model and not an explicit specification such as in the minority game.

\begin{figure}[t]
\begin{center}
\includegraphics[width=0.48\textwidth]{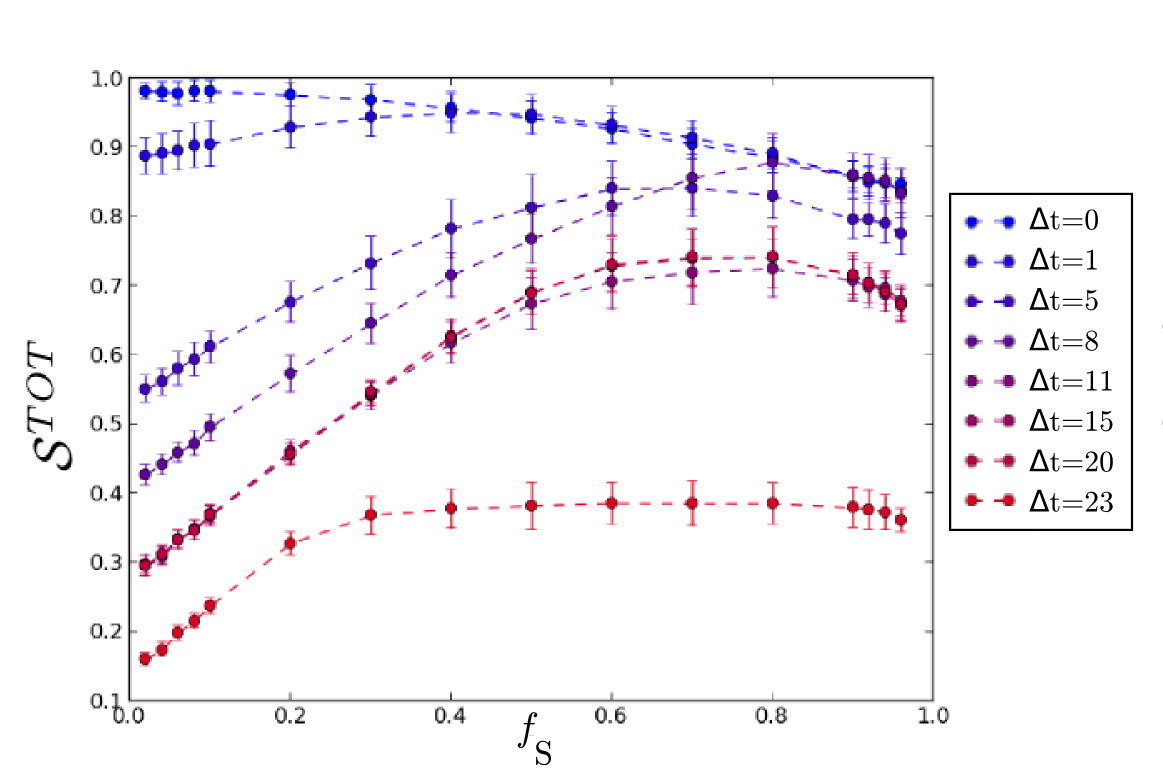}
\includegraphics[width=0.45\textwidth]{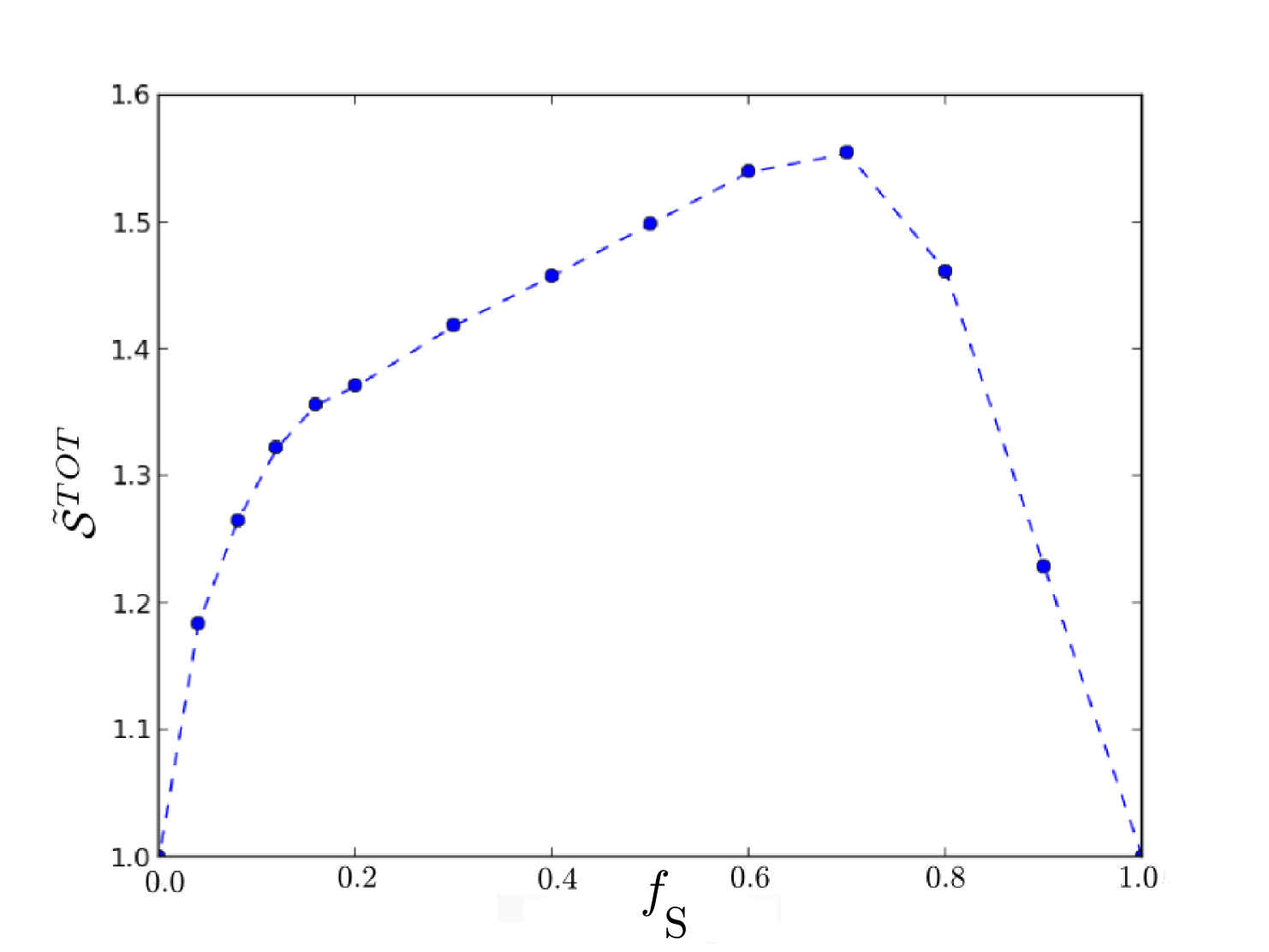}
\end{center}
\caption{Left panel. Global satisfaction against the fraction of company S for different values of $\Delta t$. The number of flights is set to 120.
Right panel. Normalized global satisfaction versus the percentage of flights R, when 100 flights are simulated.}
\label{fig:mixed_3}
\end{figure}

Finally, we consider the global satisfaction of the system, as described by ${\cal S}^{TOT}$ in Eq. \ref{eq:GS}. Left panel of figure \ref{fig:mixed_3} shows the dependence of global satisfaction from the fraction $f_S$ of S flights. It is interesting to note how it is different from Figure \ref{fig:pure_1} (right). In the former we have mixed populations with extreme values of the cost parameters, while in the latter we have pure populations with intermediate values of the cost parameters. The comparison indicates that it is not trivial to find a pure population with a behavior similar to the mixing of two extreme companies. In particular, the satisfaction is hardly monotonic now, except for very small values of $\Delta t$. We note that the overall tendency is that a uniform distribution of departing time increases the satisfaction, except for very pure populations, as we saw before. Moreover, even a high value of $\Delta t$ does not strongly favor S companies. In fact, the plot is almost entirely flat for $\Delta t=23$ and thus the result is insensitive to the fraction of S or R companies. 

As already mentioned, the $\tilde {\cal S}^{TOT}$ is meant to be insensitive to the difference of satisfaction between the pure populations, focusing instead on the interaction between the different strategies. The plot for $\tilde {\cal S}^{TOT}$  (right panel of figure \ref{fig:mixed_3}) shows that there is a positive interaction between companies, i.e. a mixture of strategies always raise the satisfaction above the base levels. Moreover, the strongest interaction occurs for a precise level of mixing (here $f_S\simeq 0.3$). Any other case is sub-optimal. Notice that the existence of the maximum is not {\it a priori} obvious, but is related to the strong positive dependence of both satisfactions with the fraction of flights. To be specific, consider a toy example in which 
\begin{equation}
{\cal S}^{(i)}=\frac{A^{(i)}}{f_i^\eta},
\end{equation}
where $i=R$ or $S$, ${\cal S}^{(i)}$ and $f_i$ are the satisfaction and fraction of flights of company $i$, $A^{(i)}$ are parameters, and $\eta>0$ describes the inverse relation between fraction and satisfaction. The $\tilde {\cal S}^{TOT}$ is
\begin{equation}
\tilde {\cal S}^{TOT}=f_S \frac{{A^{(S)}}{f_S^\eta}}{A^{(S)}}+f_R \frac{{A^{(R)}}{f_R^\eta}}{A^{(R)}}=f_S^{1-\eta}+(1-f_S)^{1-\eta}
\end{equation}
because $f_R=1-f_S$. If $\eta=1$, this function is a constant, if $\eta>1$ the function $\tilde {\cal S}^{TOT}$ has a minimum in $f_S=0.5$, while if $\eta<1$, $\tilde {\cal S}^{TOT}$ has a maximum in $f_S=0.5$. Thus the observation of left panel of figure \ref{fig:mixed_3} is not trivial and depends on the functional dependence of the satisfaction of a company from the its fraction of flights. In conclusion, when one takes into account the specificity of the cost function parameters of each company, a mixed solution is optimal with respect to a pure case.

\subsection{The role of the topology of the sector network}\label{sec:topology}

All the above results have been obtained with the same pair of airports on the same network, only changing the parameters of the model. In this section, we briefly investigate what happens in the generic case. How do the above results depend on the topology of the network connecting the two airports? In particular one can expect that the presence of several alternative and non overlapping sector paths connecting the two airports could improve the advantage of R companies.   

We built a new network of 1,000 nodes (compared to 90 previously used) to avoid border effects. In this network, we have access to more than 15,000 pairs of airports lying at the same topological distance as in the previous analyses (6 edges). Among these, we select 1,000 of these pairs for the analysis. Then we perform simulations on them independently.

One of the main points explained above is that a well spaced departing time wave structure gives an advantage to company S over company R, whereas a more uniform departing time pattern gives an advantage to R company. This advantage can be quantified by the  difference 
\begin{equation}\label{advantage}
\delta {\cal S} (\Delta t)= {\cal S}^{(S)} (\Delta t) - {\cal S}^{(R)} (\Delta t) 
\end{equation}
where ${\cal S}^{(S)} (\Delta t)$ (${\cal S}^{(R)} (\Delta t) $) is the average satisfaction of company S (R) when the departing time pattern has waves separated by $\Delta t$.
Performing simulations on the thousand pairs independently, we first checked the sign of $\delta {\cal S} (\Delta t)$ in the case where $\Delta t=23$ , i.e. where there is only one peak. We found that only very few cases give a negative value of $\delta {\cal S}(23)$. Hence, company S has always the advantage when there is only one peak. Given that this value has always the same sign, we focus on the other extreme, $\delta {\cal S} (0)$, corresponding to a uniform departing time pattern, to see if the company R has always a higher satisfaction in this case. 

\begin{figure}[t]
\begin{center}
\includegraphics[width=0.48\textwidth]{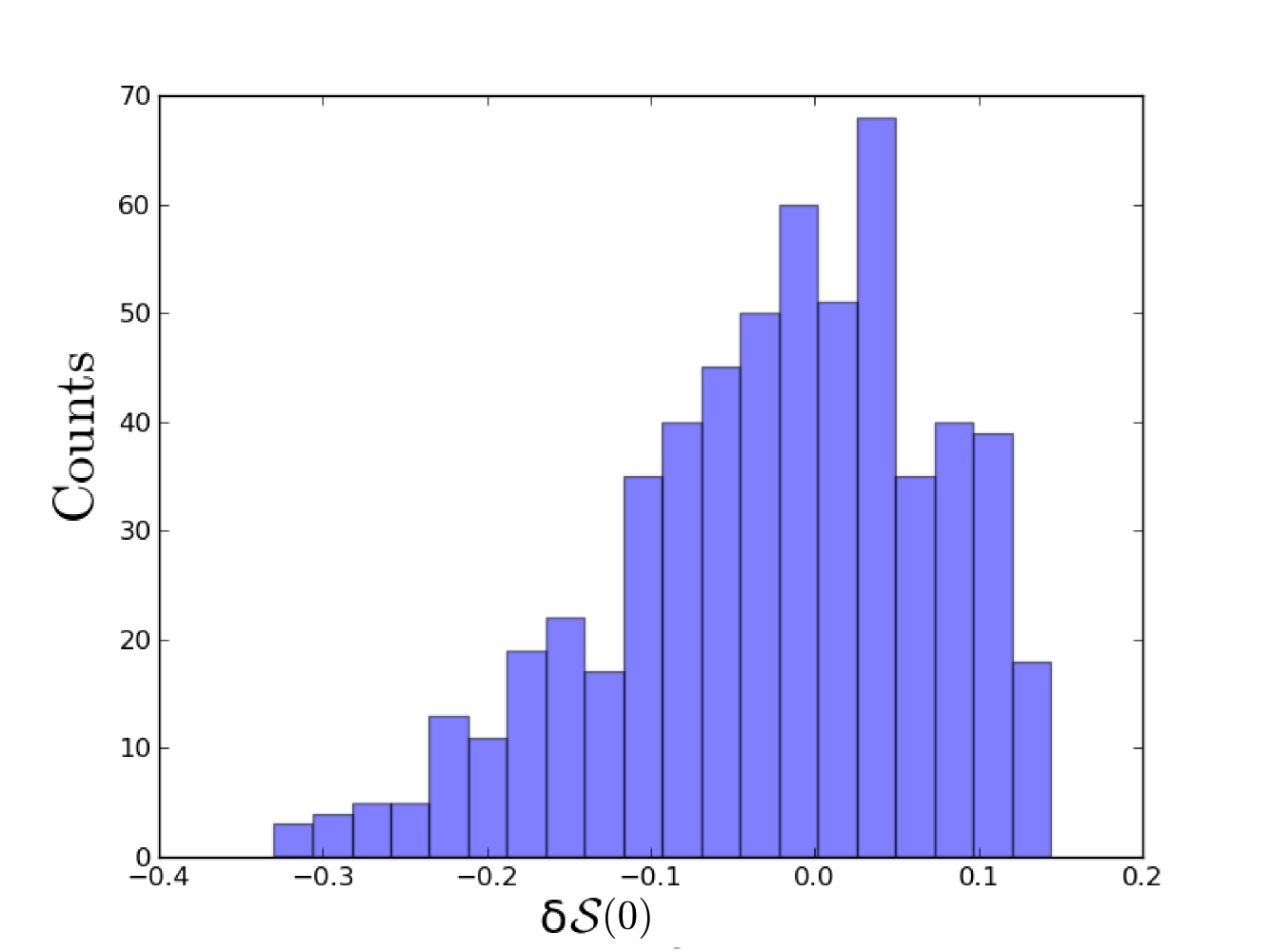}
\includegraphics[width=0.48\textwidth]{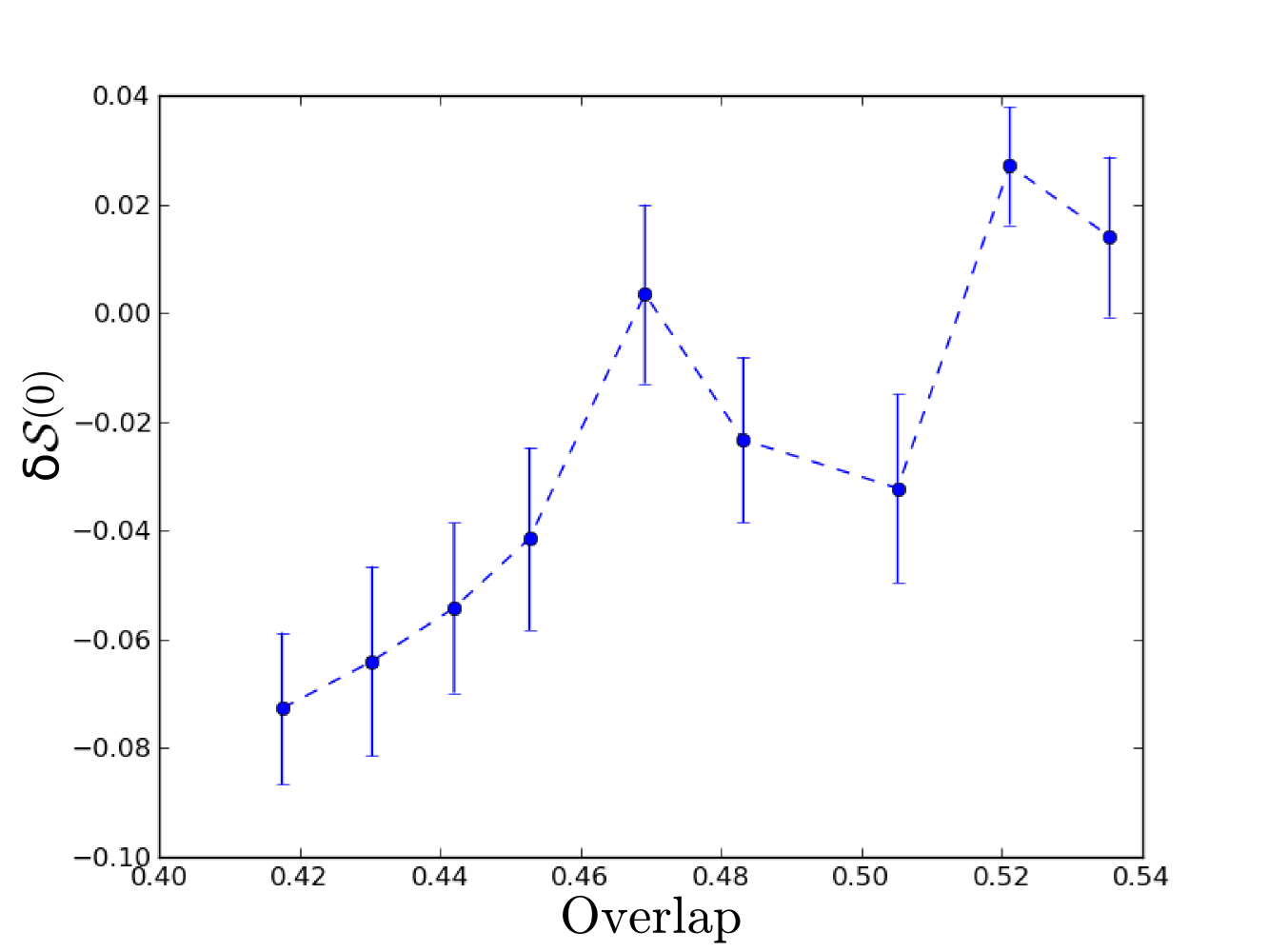}
\end{center}
\caption{Left. Histogram of the quantity $\delta {\cal S}(0)$ (see Eq. \ref{advantage}) giving the difference in satisfaction between company S and R when the departing time pattern is uniform ($\Delta t=0$). Right. Relationship between overlap and $\delta {\cal S}(0)$ (see Eq. \ref{advantage}), giving the difference in satisfaction between company S and R when the departing time pattern is uniform  (i.e. $\Delta t=0$). We show the average on each quantile of overlap. Error bars are standard deviations.}
\label{fig:hist_swap}
\end{figure}

Figure \ref{fig:hist_swap} (left panel) shows the histogram of the quantity  $\delta {\cal S}(0)$. As we can see, there is a sizable, yet smaller than 50\%, fraction of cases where $\delta {\cal S}(0)>0$. This means that in these cases, company S is still getting a larger satisfaction than company R. What are these cases? Intuitively, the R company does not get any advantage when the rerouted flights keeps hitting at least one overloaded sector. This means that when a rerouted flight plan is crossing a sector which is already present in a previously rejected flight plan, it will be rejected by the NM. Hence, an important measure here is the overlap between the possible paths connecting the two airports. Our hypothesis is that if this quantity is large, there is no advantage to reroute the flight, even when a uniform departing time pattern would give an advantage to company R with respect to S.

To test this hypothesis, given the set $P_{AB}$ of the $N_{fp}$ shortest paths between a given couple of airports $A$ and $B$, we define the overlap $O(P_{AB})$ as 
\begin{equation}
O(P_{AB}) = \frac{1}{N_{fp}(N_{fp} - 1)} \sum_{{ \bf p_1}\in P_{AB}}\sum_{{ \bf p_2}\in P_{AB}} \frac{\# \lbrace{\mathcal N}({\bf p_1}) \cap  {\mathcal N}(\bf{p_2})\rbrace}{\#\lbrace {\mathcal N}({\bf p_1}) \cup  {\mathcal N}(\bf{p_2})\rbrace},
\end{equation}
where ${\mathcal N}(\bf{p})$ is the set of sectors in the path $\bf{p}$ and $\#$ indicates the cardinality of the set. The overlap $O(P_{AB})$ is the ratio of the number of sectors which are common to two paths over the total number of (unique) sectors in these paths, averaged over all the pairs of paths connecting airports $A$ and $B$. Note that we have included both airports in the metric, so $O(P_{AB})$ cannot be equal to 0. It cannot reach 1 either, since it would mean that every path is the same\footnote{More precisely, it means that they are composed of the same sectors, which could in principle be arranged in different orders. This is very unlikely to happen for $N_{fp}=10$.}. In figure \ref{fig:hist_swap} (right panel) we show the relationship between this overlap, computed for each pair of airport, and $\delta {\mathcal S}(0)$. There is a clear positive correlation between the two quantities. In particular it shows that the bigger is the overlap between the paths connecting the two airports, the bigger the advantage of S over R. This finding sheds light on our previous results. The mechanisms we have found are present in the system, but are potentially hidden by some special configuration of the network for certain pair of airports. For the two airports considered in the previous sections, the overlap is $0.48$, giving a negative value of $\delta {\mathcal S}(0)$, consistently with the fact that uniform departing time gives an advantage to R companies over S. 

In conclusion, the precise interplay between the environment (the network), the common trends (the waves), and the individual behaviors (the parameters of the cost function) gives rise to different advantage for each agent, and thus to different advantage, as measured by satisfaction, of different companies.

\section{Conclusions}\label{sec:conclusions}

We presented in this paper a toy model for the allocation of flight plans in an airspace. In short, air companies build an optimal flight plan and submit it to a network manager, which allocates it to the airspace if it is possible, i.e. if it is not already congested. May the flight plan be rejected, the air companies submit another suboptimal flight plan, which will differ from the first one depending on their idiosyncratic cost function. The satisfaction of the air company is then related to the number of flight plans rejected and the ratio between the cost of the accepted flight plan and the desired one.

We first studied the case of a given pair of airports and showed that there is a (congestion) transition, much like the congestion observed in other transport systems, e.g. car traffic. By considering a whole range of strategies, we showed that their respective advantage depends on their environment, here mainly the pattern of times of departure. In particular, intermediate strategies can sometimes outperform extreme strategies, even in a pure population setting.

By mixing two extreme types of companies, we showed that the system exhibits some strong analogies with theoretical games like hawks and doves or the Minority Game. In particular, we showed that there exists a unique fraction of mixing corresponding to an evolutionary stable strategy or a stable Nash equilibrium, depending on the point of view. We also found that the strategies are interacting positively, leading to an absolute maximum in satisfaction for the overall system at a mixing fraction different from 0 and 1.

Finally, we checked the general validity of the previous results, only performed on a small network and a fixed pair of airports. We found that the general case is consistent with this particular one. However, we showed that all the previous effects can be hidden by some particular configurations of the airspace. In particular, the possibility to gain advantage of a rerouting is closely related to the overlap between the different possible paths connecting two airports.

This last result in particular gives the opportunity to reflect on the importance of policy. As already mentioned in introduction, policy-maker have usually little control over the strategies involved but can act on the infrastructure itself, i.e. the network and, partly, the departing times. Hence, we foresee some possible improvements in the traffic by altering the topology of the airspace to allow more diversity in the strategies, or give advantage to a particular kind of strategy. 

It is interesting to note that the structure of the presented model could be used in other networked systems. For example in a communication network, where nodes have a finite transmission capacity, the choice between rerouting and delaying the message is equivalent to the way in which we model the major airline carriers, who prefer to reroute but to be on time, and the low cost airlines, which prefer to delay the flight but use the shortest path.

\section*{Acknowledgments}

We thanks Marc Bourgois, Salvatore Miccich\'e, and Simone Pozzi for useful discussions. 
The work presented therein was co-financed by EUROCONTROL on behalf of the SESAR Joint Undertaking in the context of SESAR Work Package E project ELSA ``Empirically grounded agent based model for the future ATM scenario''. The paper reflects only the authors' views.  EUROCONTROL is not liable for any use that may be made of the information contained therein.

  \end{document}